\documentclass[12pt]{article} 
  
\usepackage{amsfonts,amssymb}  
\usepackage{theorem}  

\def\G{\Gamma}    
\def\GG{{\rm I}\!\Gamma}   
\def\h{\tilde h}
\def\C{\tilde C}
\def\omegaT{\tilde \omega}
\def\psiT{\tilde\psi}
\def\xiT{{\tilde\xi}}
\def\thetaT{{\tilde\theta}}
\def\barthetaT{{\tilde {\bar \theta}}}
\def\baromegaT{{\tilde {\bar \omega}}}

\begin{document}

\rightline{MPI-Pht-2003-29}
\rightline{MPP-2003-43}

\vskip 2.8 truecm
\Large
\bf
\centerline{On a class of embeddings}
\centerline{of massive Yang-Mills theory}

\vskip 1.3 truecm
\rm
\normalsize
\begin{center}
Andrea Quadri\\ 
Max-Planck-Institute for Physics\\  
(Werner-Heisenberg-Institute) \\ 
F\"ohringer Ring 6, D-80805, Munich, Germany
\end{center}

\vskip 1.5 truecm
\begin{quotation}
A power-counting renormalizable model into which massive
Yang-Mills theory is embedded is analyzed. 
The model is invariant under a nilpotent BRST
differential $s$. The physical observables of the embedding theory,
defined by the cohomology classes of $s$ 
in the Faddeev-Popov neutral sector, are given
by local gauge-invariant quantities constructed only from the field
strength and its covariant derivatives.
\end{quotation}


\newpage

\section{Introduction}

In the search for a renormalizable model of massive
Yang-Mills theory it has become apparent since a long time
that at least one additional scalar field is required
in order to fulfill physical unitarity already at tree level
\cite{'tHooft:rn,Cornwall:1973tb,Cornwall:1974km,LlewellynSmith:ey,Martin}.
 
The spontaneous breaking of gauge symmetry provides a way
to account for massive non-Abelian gauge bosons within
the class of power-counting renormalizable models
by means of the Higgs mechanism (see e.g. \cite{weinberg}).
In its simplest version one additional physical scalar
particle is introduced in the spectrum.
The corresponding field acquires a non-vanishing
vacuum expectation value ({\it v.e.v.}) due to the presence
of a suitably designed quartic potential, which triggers
the spontaneous gauge symmetry breaking.

On the other hand, in the BRST quantization of gauge theories \cite{BRST,
Becchi:1975nq,Becchi:bd}
the requirement of 
physical unitarity is translated into the nilpotency of the relevant
BRST differential, together with the condition of the
absence of quantum anomalies for 
the corresponding Slavnov-Taylor (ST) identities
\cite{Becchi:1974xu,Curci:1976yb,Kugo:zq}.
Both requirements are fulfilled in the models based on the Higgs mechanism
\cite{Becchi:jx}.

\medskip
In this paper we would like to address an alternative method
to generate a mass for non-Abelian gauge bosons, based on a 
fixing condition for an additional shift symmetry of a larger model,
into which massive Yang-Mills theory can be embedded.

We wish to study whether it is possible to construct
an embedding theory, with a suitably chosen field content,
which is invariant under a nilpotent BRST differential $s$ such that
all the fields, besides the gauge fields $A_\mu^a$ and
their gauge ghost fields $\omega^a$, pair into BRST doublets
with their shift ghost partners.
As a consequence, these additional fields do not affect
the cohomology classes of $s$, which reduce in the
sector with zero Faddeev-Popov (FP) charge (physical observables)
to gauge-invariant quantities constructed only from the field
strength and its covariant derivatives.

\medskip
A first possibility to explore in order to find the correct
embedding model for massive Yang-Mills theory 
is to
apply the embedding procedure to the field content of the ordinary 
Higgs model in the presence of a single Higgs multiplet.
As we will show, the extension necessarily contains 
in the bosonic sector a scalar field $\phi$  and a
vector field $b_\mu$. 
$\phi$ is to be identified with the scalar field appearing in the
usual Higgs mechanism. The component of $\phi$ which acquires
a non-zero {\em v.e.v.} is in correspondence with the standard Higgs field.

The resulting model  turns out
to have good UV properties,
but it shows a violation
of the IR power-counting conditions which destroys
renormalizability by power-counting.
%
As a consequence, the embedding
procedure in the presence of the Higgs mechanism fails to produce
the sought-for renormalizable 
embedding of massive Yang-Mills theory.
Nevertheless, from the analysis of the Higgs
model embedding it can be readily seen, on the basis
of cohomological arguments,
that a different field content leads to a possible successful embedding
of massive Yang-Mills theory.
In this minimal embedding model the field $\phi$ is no longer
needed and only the additional field $b_\mu$
appears in the bosonic sector together with the gauge fields $A_\mu^a$.

\medskip
The minimal embedding model fulfills both the UV and the IR
power-counting conditions and is therefore power-counting renormalizable.
It also shows a simple cohomological structure in the extended
field sector.

The most important feature of the minimal embedding model is 
that spontaneous gauge symmetry breaking is not realized in it
and still the physical spectrum contains massive gauge fields. 
The set of the physical observables of the theory is given
by the local gauge-invariant quantities,
constructed only from the field
strength and its covariant derivatives.
Hence the minimal embedding model could be considered as 
a candidate for a renormalizable
theory of massive non-Abelian gauge bosons.

\medskip
At the diagrammatic level it can be seen that the physical
scalar required in the unitarity diagrams 
\cite{Lee:1977eg} is now provided by $\partial b$.
It should be noted however that
 all the properties of the minimal embedding model are actually a direct and natural 
consequence of its BRST symmetry and can be better
understood on purely cohomological grounds.
One can show that the relevant ST identities 
are anomaly-free and therefore they can be restored order by order
in perturbation theory by a suitable choice
of finite action-like counterterms 
\cite{Piguet:er,Ferrari:1999nj,Quadri:2003ui,Quadri:2003pq}.
This in turn guarantees the consistent quantization of the model
to all orders in the loop expansion.

In this paper we focus on 
renormalizability and on the off-shell 
cohomological properties of the theory, with the aim
of identifying the local physical observables.
Moreover we also propose a method to characterize the relevant asymptotic 
states, by making use of techniques developed to study
the on-shell cohomology of gauge theories \cite{Becchi:bd,Curci:1976yb}.

Many more things remain to be done. The observables of the model
could be computed in perturbation theory and compared with the corresponding
quantities in other models
for massive gauge bosons. The construction proposed in this paper
might also suggest different alternatives in order to obtain 
a candidate for massive Yang-Mills theory. The question of the 
inclusion of massive fermions within the embedding formalism
should also be analyzed, in view of possible phenomenological applications
of this class of theories.

\medskip

The paper is organized as follows. 
In Sect.~\ref{sect1} we analyze the embedding of the ordinary
Higgs model with a single Higgs multiplet 
and discuss the associated embedding BRST differential.
We study the UV and IR properties of the resulting theory and
show that it fails to be renormalizable by power-counting,
due to a bad behaviour in the IR regime. 

In Sect.~\ref{minim} we move to the analysis of the minimal
embedding procedure and show that it leads to a 
well-defined power-counting renormalizable
theory. For definiteness we work with the gauge group
$SU(2)$, but the construction can be straightforwardly
extended to other groups of interest, including those
involving Abelian factors like $SU(2) \times U(1)$.
The physical observables of this model,
as selected by the cohomology of the relevant BRST
differential $s$ in the FP-neutral sector,
are the gauge-invariant quantities constructed
from the field strength and its covariant derivatives.
Moreover, we analyze 
the structure of the asymptotic states and show that
they contain three massive $SU(2)$ gauge bosons.
Their physical polarizations are selected by the linearized
BRST differential $\hat s$ to be the expected transverse ones.
Finally conclusions are given in Sect.~\ref{sect4}.

\vskip 0.1 truecm

Appendices \ref{appA} and \ref{appC} contain the 
detailed construction of the Hilbert spaces needed in order 
to identify
the physical asymptotic states of the minimal embedding model,
while
the UV and IR degrees of the fields are summarized in Appendix \ref{appB}.

\section{The embedding procedure in the presence of the Higgs mechanism}
\label{sect1}

As pointed out in the introduction,
an extended set of fields which pair into BRST doublets
is needed in order to construct an embedding model 
for massive Yang-Mills theory, 
while guaranteeing the nilpotency of the relevant BRST differential.

Since the new fields will form BRST doublets,
they do not affect the cohomology of the model 
\cite{Gomis:1994he,Piguet:er,Barnich:2000zw}.
As a consequence, the physical observables of the embedding theory,
defined by the cohomology classes of the relevant BRST
differential in the FP-neutral sector, coincide with the
gauge-invariant quantities constructed only from the field
strength and its covariant derivatives.

\medskip
A first possibility to explore in order to find the correct
embedding model for massive Yang-Mills theory is to
apply this procedure to the field content of the ordinary 
Higgs model in the presence of a single Higgs multiplet.
If power-counting renormalizability is dropped, 
the St\"uckelberg model \cite{Dragon:1996tk}
and its generalizations \cite{Banerjee:1997sf}
could also be considered. However, in this paper we aim at
finding a power-counting renormalizable embedding for
massive Yang-Mills theory, and we will not discuss
such alternatives. 

As we will show, the extension necessarily contains 
in the bosonic sector a scalar field $\phi$  and a
vector field $b_\mu$. 
$\phi$ is to be identified with the scalar field appearing in the
usual Higgs mechanism. The component of $\phi$ which acquires
a non-zero {\em v.e.v.} is in correspondence with the standard Higgs field.

\medskip
In the embedding of the Higgs model we require
that  the complex scalar $\phi$ forms a
doublet pair together with its BRST partner $h$.
We denote by $T^a$ the generators of the Lie algebra $\mathfrak{g}$
acting on the representation to which $\phi,h$ belong.
The commutation relations are
\begin{eqnarray}
[T^a,T^b] = i f^{abc} T^c \, .
\label{e1}
\end{eqnarray}
The normalization is chosen in such a way that
\begin{eqnarray}
Tr (T^aT^b) = 2 \delta^{ab} \, .
\label{e1_bis}
\end{eqnarray}
The covariant derivative on $\phi$ is
\begin{eqnarray}
D_\mu \phi(x) = \partial_\mu \phi(x) - i A_{\mu a} T^a \phi(x)
\label{e2}
\end{eqnarray}
and analogously for $h$.
The BRST differential $s$ acts on $\phi,h$ as follows
\footnote{Strictly speaking, in eq.(\ref{e3}) the BRST doublet
is formed by $(\phi, i \omega_a T^a \phi + h)$. We will comment
later on on the decomposition of the full BRST differential $s$
into its gauge part and its pure shift part.}
\begin{eqnarray}
&& s \phi = i \omega_a T^a \phi + h \, , ~~~~
s \phi^\dagger = - i \omega_a \phi^\dagger T^a  + h^\dagger \, , 
\nonumber \\
&& s h = i \omega_a T^a h \, , ~~~~
   s h^\dagger = - i \omega_a  h^\dagger T^a \, .
\label{e3}
\end{eqnarray}
$\omega^a$ are the ghost fields associated to the gauge
transformation of the gauge group $G$ whose Lie algebra
is $\mathfrak{g}$.
The BRST transformation of the gauge field $A_\mu^a$ is
\begin{eqnarray}
s A_\mu^a = (D_\mu \omega)^a = \partial_\mu \omega^a 
+ f^{abc} A_\mu^b \omega^c \, ,
\label{e3Bis}
\end{eqnarray}
while the BRST transformation of the ghost fields is
\begin{eqnarray}
s \omega^a = -\frac{1}{2} f^{abc} \omega^b \omega^c \, .
\label{e3Ter}
\end{eqnarray}
$\phi,\phi^\dagger$ are commuting, $h,h^\dagger$ and $\omega^a$
are anticommuting.

The Faddeev-Popov (FP) charge is assigned as follows:
\begin{eqnarray}
{\rm FP}(\phi) = {\rm FP}(\phi^\dagger) = 0 \, , ~~~~
{\rm FP}(h) = {\rm FP}(h^\dagger) = 1 \, .
\label{e4}
\end{eqnarray}
The ghost fields $\omega^a$ have FP charge $+1$.

\medskip
Since $\phi$ is a component of a BRST doublet
it follows from general cohomological arguments  
\cite{Gomis:1994he,Piguet:er,Barnich:2000zw}
that in any action invariant
under the BRST differential $s$ all terms depending on $\phi$
must be $s$-exact. Since we wish to include in the classical
action of the embedding of the Higgs model the term
$(D_\mu \phi)^\dagger D^\mu \phi$, we need 
in the sector with FP-charge $-1$ an anticommunting
antighost field $\psi^\mu$ in the same representation as $\phi$.

This also implies that a new ghost field $\xi_\mu$, 
pairing with the antifield $\psi_\mu$, will enter into the field
content of the embedding model.
$\xi_\mu$ is in the same representation as $\phi$ and has 
FP-charge $+1$. It forms a BRST doublet with a FP-neutral
field $b_\mu$, again in the same representation as $\phi$.

\medskip
We assign the following transformation rules for 
$\psi^\mu$:\footnote{We work in the on-shell formalism, avoiding the 
introduction of the Nakanishi-Lautrup multiplier fields.}
\begin{eqnarray}
&& s \psi^\mu = + i \omega_a T^a \psi^\mu + D^\mu \phi + D_\rho b^{\rho\mu} 
 \, ,
\nonumber \\
&& s \psi^{\mu \dagger} = -i \omega_a \psi^{\mu\dagger} T^a 
- (D^\mu \phi)^\dagger - (D_\rho b^{\rho\mu})^\dagger
\, .
\label{e5}
\end{eqnarray}
In the above equation we have set
\begin{eqnarray}
b_{\rho\mu} = D_{[\rho} b_{\mu]} \, ,
\label{new1}
\end{eqnarray}
where $b_\mu$ is a complex field
whose BRST transformation is given by
\begin{eqnarray}
s b_{\mu} = + i \omega_a T^a b_\mu  + \xi_\mu \, , 
~~~~ 
s b^\dagger_{\mu} =  - i \omega_a b^\dagger_{\mu} T^a + \xi^\dagger_{\mu} \, .
\label{e6}
\end{eqnarray}
In eq.(\ref{new1}) $[...]$ means antisymmetrization with respect
to the indices in the square bracket.
$\xi_{\mu}$ and $\xi_{\mu}^\dagger$ are anti-commuting
fields with FP-charge $+1$ whose
BRST transformation is
\begin{eqnarray}
s \xi_{\mu} = i \omega_a T^a \xi_{\mu} \, , ~~~~~
s \xi^\dagger_{\mu} = -i \omega^a \xi^\dagger_{\mu} T^a \, .
\label{e7}
\end{eqnarray}
%
As anticipated, the ghost fields $\xi_\mu, \xi^\dagger_\mu$ pair 
under the BRST differential $s$ with the fields $b_\mu,b^\dagger_\mu$.

\medskip
The ghost fields $\xi_\mu, \xi^\dagger_\mu$ in the sector
with FP-charge +1 are in correspondence with the antighost
fields $\psi_\mu,\psi^\dagger_\mu$ in the sector with FP-charge $-1$.
In order to preserve the correspondence between the sectors
with negative and positive FP charge 
we need to introduce the field $\zeta$ and its conjugate 
$\zeta^\dagger$, with the following BRST transformation:
\begin{eqnarray}
s \zeta = +i \omega_a T^a \zeta + \beta D^2 D^{\rho} b_\rho \, , ~~~~ 
s \zeta^\dagger = -i \omega_a \zeta^\dagger T^a - 
\beta (D^2 (D^{\rho} b_\rho))^\dagger \, .
\label{e8}
\end{eqnarray}
$\zeta,\zeta^\dagger$ have FP-charge $-1$. $\zeta$ is in the same
representation as $\phi$. $\beta$ is a dimensionless 
parameter. We will comment on its r\^ole later on 
in Sect.~\ref{sgsb}.

The conjugation rules for the BRST differential $s$ are the usual
ones: for $f$ fermion we have
\begin{eqnarray}
s f^\dagger = - (sf)^\dagger \, , 
\label{new9}
\end{eqnarray}
while for $b$ boson
\begin{eqnarray}
s b^\dagger = (sb)^\dagger \, .
\label{e9}
\end{eqnarray}

\medskip
The action
\begin{eqnarray}
S & = &  
Tr \int d^4x \, \left [ (D_\mu \phi + D^\nu b_{\nu\mu})^\dagger
                        (D^\mu \phi + D_\rho b^{\rho\mu})
\right .
\nonumber \\
& & ~~~~~~~~~~~~~ \left .
- \beta^2 (D^\rho b_\rho)^\dagger D^2 (D^{\rho'} b_{\rho'})
\right . \nonumber \\
& & ~~~~~~~~~~~~~ \left . 
+ \psi_\mu^\dagger ( D^\mu h + D_\rho D^{[\rho} \xi^{\mu]})
               - \psi_\mu (D^\mu h + D_\rho D^{[\rho} \xi^{\mu]})^\dagger
\right . \nonumber \\
& & ~~~~~~~~~~~~~ \left .
- \beta \zeta^\dagger D^\rho \xi_\rho + \beta \zeta 
 (D^\rho \xi_\rho)^\dagger \right ] \nonumber \\
& = & 
Tr \int d^4x \, \left [ (D_\mu \phi)^\dagger D^\mu \phi
		       - b^\dagger_{\mu\nu} D^\mu D^\nu \phi -
		        (D^\mu D^\nu \phi)^\dagger b_{\mu\nu} 
		      -b^\dagger_{\mu\nu} D^\mu D_\rho b^{\rho\nu} \right .
\nonumber \\
& & ~~~~~~~~~~~~~ \left .
- \beta^2 (D^\rho b_\rho)^\dagger D^2 (D^{\rho'} b_{\rho'})
\right . \nonumber \\
& & ~~~~~~~~~~~~~ \left . 
+ \psi_\mu^\dagger ( D^\mu h + D_\rho D^{[\rho} \xi^{\mu]})
               - \psi_\mu (D^\mu h + D_\rho D^{[\rho} \xi^{\mu]})^\dagger
\right . \nonumber \\
& & ~~~~~~~~~~~~~ \left .
- \beta \zeta^\dagger D^\rho \xi_\rho + \beta \zeta 
 (D^\rho \xi_\rho)^\dagger \right ]
\label{e10}
\end{eqnarray}
is BRST-invariant. $s$ is nilpotent except on $\psi_\mu, \psi_\mu^\dagger$,
on which it squares to zero modulo the equations of motion of
$\psi_\mu^\dagger, \psi_\mu$:
\begin{eqnarray}
s^2 \psi_\mu =  \frac{\delta S}{\delta \psi_\mu^\dagger} \, , ~~~~
s^2 \psi_\mu^\dagger =  \frac{\delta S}{\delta \psi_\mu} \, ,
\label{e11}
\end{eqnarray}
and on $\zeta,\zeta^\dagger$, on which we get
\begin{eqnarray}
s^2 \zeta = - D^2 \frac{\delta S}{\delta \zeta^\dagger} \, , ~~~~
s^2 \zeta^\dagger = - \frac{\delta S}{\delta \zeta} {\overleftarrow{D}^2}^\dagger \, .
\label{e11bis}
\end{eqnarray}

We observe that $s$ admits the following decomposition:
\begin{eqnarray}
s = s_0 + s_1 \, ,
\label{e12} 
\end{eqnarray}
with
\begin{eqnarray}
&& s_0 A_\mu^a = (D_\mu \omega)^a \, , ~~~~ 
s_0 \omega^a = -\frac{1}{2} f^{abc} \omega^b \omega^c \, , \nonumber \\
&&
s_0 \Phi = i \omega_a T^a \Phi \, , ~~~~ s_0 \Phi^\dagger = -i \omega_a \Phi^\dagger T^a \, , 
\label{e12Bis}
\end{eqnarray}
where $\Phi$ is a collective notation for $\{ \phi, b_{\mu}, \psi_\mu, 
\xi_{\mu},h, \zeta \}$, and
\begin{eqnarray}
&& s_1 \phi = h \, , ~~~~~~ s_1 h = 0 \, , ~~~~~
   s_1 \phi^\dagger = h^\dagger \, , ~~~~ s_1 h^\dagger = 0 \, ,
   \nonumber \\
&& s_1 b_{\mu} = \xi_{\mu} \, , ~~~~ s_1 \xi_{\mu} = 0 \, ,
   ~~~~ s_1 b^\dagger_{\mu} = \xi^\dagger_{\mu} \, , ~~~~
   s_1 \xi^\dagger_{\mu} = 0 \, , 
   \nonumber \\
&& s_1 \psi_\mu = D_\mu \phi + D^\rho D_{[\rho} b_{\mu]} \, , ~~~~
   s_1 \psi_\mu^\dagger = - (D_\mu \phi + D^\rho D_{[\rho} b_{\mu]})^\dagger \, ,
\nonumber \\
&& s_1 \zeta = \beta D^2 D^\rho b_\rho \, , ~~~~
   s_1 \zeta^\dagger = - \beta ( D^2 (D^\rho b_\rho))^\dagger \, , 
\nonumber \\
&& s_1 A_\mu^a = 0 \, , ~~~~ s_1 \omega_a = 0 \, .
\label{e13}
\end{eqnarray}
$s_0$ in eq.(\ref{e12}) implements the gauge BRST transformations,
$s_1$ the shift BRST transformations.

$S$ in eq.(\ref{e10}) is simultaneously $s_0$- and $s_1$-invariant.
Therefore it is also invariant under the following differential
\begin{eqnarray}
s_\lambda = s_0 + \lambda s_1 \, ,
\label{e14}
\end{eqnarray}
where $\lambda$ is a commuting constant with FP-charge zero.

We now consider the following $s_\lambda$-invariant action
\begin{eqnarray}
S_{embed} = -  \frac{1}{4g^2} \, \int d^4x \, F_{\mu\nu}^a F^{\mu\nu \, a} + S \, ,
\label{e15}
\end{eqnarray}
where the field strength $F_{\mu\nu}^a$ is given by
\begin{eqnarray}
F_{\mu\nu}^a = \partial_\mu A^a_\nu - \partial_\nu A_\mu^a 
	      +f^{abc} A_\mu^b A_\nu^c \, .
\label{e15bis}
\end{eqnarray}
$g$ is the coupling constant of the group $G$.
The assignment of mass dimension of the fields is as follows:
$A_\mu^a,\omega^a$,$\phi,\psi_\mu,\xi_\mu$ and the
conjugates $\phi^\dagger,\psi_\mu^\dagger,\xi_\mu^\dagger$
have dimension $1$, $\zeta,h$ and the conjugate
$\zeta^\dagger,h^\dagger$ have dimension $2$ while 
$b_\mu$ and $b^\dagger_\mu$ have dimension zero.

\medskip
$S$ in eq.(\ref{e10}) has to be considered as a fixing
functional for the shift symmetry associated with the
BRST differential $s_1$. On the other hand
the gauge invariance associated with the differential $s_0$
is still preserved by the full $S_{embed}$  in eq.(\ref{e15}).

\subsection{Spontaneous gauge symmetry breaking in the Higgs embedding model}
\label{sgsb}

The equations of motion for $\phi,\phi^\dagger$, derived
from the action in eq.(\ref{e15}), admit
the constant solution
\begin{eqnarray}
\phi= \phi_v \, , ~~~~ \phi^\dagger = \phi_v^\dagger \, , ~~~~~
\partial_\mu \phi_v = 0 \, ,
\label{gm1}
\end{eqnarray}
with all other fields equal to zero.
The shift 
\begin{eqnarray} 
\phi \rightarrow \phi_v + \phi \, , ~~~~ 
\phi^\dagger \rightarrow \phi^\dagger_v+  \phi^\dagger
\label{gm2}
\end{eqnarray}
yields for the action $S_{embed}$
\begin{eqnarray}
S_{embed} & = & 
            - \frac{1}{4g^2} \int d^4x \, F_{\mu\nu}^a F_{\mu\nu a}
            + S \nonumber \\
	  & & + \int d^4x \, A_{\mu a} M^{ab} A^{\mu}_b \nonumber \\
	  & & + Tr \int d^4x \, \left ( i A_{\mu a} \phi_v^\dagger T^a D^\mu \phi + h.c. \right )
              \nonumber \\
	  & & + Tr \int d^4x \, \left ( \frac{i}{2} b^{\mu\nu\dagger} F^a_{\mu\nu} T^a \phi_v + h.c. \right )
\label{em1}             
\end{eqnarray}
where
\begin{eqnarray}
M^{ab} = Tr [ \phi_v^\dagger T^a T^b \phi_v] \, .
\label{em2}
\end{eqnarray}
The term in the second line 
in eq.(\ref{em1}) is a mass term for the non-Abelian gauge fields
of the same kind as in the usual Higgs mechanism.

\medskip

Some comments are in order here. Due to the shift symmetry $s_1$ it is possible
to introduce the term $(D_\mu \phi)^\dagger (D^\mu \phi)$ only via 
the functional in the first line of eq.(\ref{e10}). 
The latter provides the fixing condition for the shift symmetry of the model.
The needed vector antighost field $\psi_\mu$ requires
the introduction of the vector ghost field $\xi_\mu$, pairing with the
vector field $b^\mu$ into a $s_1$-doublet. 
The antighost field $\zeta$ is associated
with the $s_1$-partner $h$ of $\phi$ 
and is needed to generate the term
$$-\beta^2 (Db)^\dagger D^2 (Db)$$
in the second line of eq.(\ref{e10}).

In the absence of such a term (i.e. at $\beta=0$) the propagator
of $b_\mu$ does not exist. 
We remark that, as a consequence of the shift symmetry $s_1$,
no quartic potential for the field $\phi$ is allowed.

\subsection{BV formulation of the Higgs model embedding}
\label{cohom_spectrum}

Since the BRST differential $s$ squares to zero only modulo the
equations of motion of $\psi_\mu,\psi^\dagger_\mu$ and of
$\zeta,\zeta^\dagger$, the Batalin-Vilkovisky (BV) formalism must be used
to derive the solution $\G^{(0)}$ to the classical ST identities
of the model \cite{Gomis:1994he}. 

$\G^{(0)}$ contains terms quadratic in the antifields
$\psi_\mu^{*\dagger}, \psi_\mu^*$ and $\zeta^{*\dagger}, \zeta^*$
and is given by
\begin{eqnarray}
\G^{(0)} & = & S_{embed} +
	        \int d^4x \,  \Bigg [ \frac{\alpha}{2}
	       \Big ( \partial A^a + \frac{1}{\alpha}
	       Tr ( i \phi_v^\dagger T^a \phi - i \phi^\dagger T^a \phi_v ) \Big )^2
\nonumber \\ 
& & - \alpha \bar \omega^a s \Big ( \partial A^a + \frac{1}{\alpha}
	       Tr ( i \phi_v^\dagger T^a \phi - i \phi^\dagger T^a \phi_v )
              \Big )  
\Bigg ]
\nonumber \\
& & + \int d^4x \, A_\mu^{a*} (D_\mu \omega)^a 
    - \int d^4x \, \omega^{a*}  \frac{1}{2} f^{abc} \omega^b \omega^c 
\nonumber \\
& & +  \,  Tr \int d^4x \,  \chi^* s \chi 
    +  \, Tr  \int d^4x \, \chi^{ *\dagger} s \chi^\dagger 
\nonumber \\
        & & + Tr \int d^4x \, \left ( - \psi_\mu^{*\dagger} \psi^{*\mu}
	                            +\zeta^* D^2 \zeta^{\dagger *} \right ) \, .
\label{em4}
\end{eqnarray}
where $S_{embed}$ is given in eq.(\ref{em1}),
$\chi$ is a collective notation for $\{ \phi,h,b_\mu,\xi_\mu,\psi_\mu,\zeta\}$
and 
$\chi^*$ stands for the antifields $\{ \phi^*,h^*,b^*_\mu, \xi^*_\mu, \psi^*_\mu, \zeta^*\}$.
We have chosen a $R_\alpha$-gauge by setting
\begin{eqnarray}
s \bar \omega^a =  \partial A^a + \frac{1}{\alpha}
	       Tr ( i \phi_v^\dagger T^a \phi - i \phi^\dagger T^a \phi_v ) \, .
\label{em5}
\end{eqnarray}
$\bar \omega^a$ is the antighost for the gauge ghost field $\omega^a$.
The R.H.S. of eq.(\ref{em5}) is linear in the quantum fields, therefore
we do not need an antifield for $\bar \omega^a$.

$\G^{(0)}$ satisfies the following Slavnov-Taylor (ST) identities:
\begin{eqnarray}
{\cal S}(\G^{(0)}) & = & \int d^4x \, \Bigg (
\frac{\delta \G^{(0)}}{\delta A_\mu^{a*}}
\frac{\delta \G^{(0)}}{\delta A^\mu_a}
+
\frac{\delta \G^{(0)}}{\delta \omega^{a*}}
\frac{\delta \G^{(0)}}{\delta \omega_a}
+ Tr \frac{\delta \G^{(0)}}{\delta \chi^*}
     \frac{\delta \G^{(0)}}{\delta \chi}
  + Tr \frac{\delta \G^{(0)}}{\delta \chi^{*\dagger}}
       \frac{\delta \G^{(0)}}{\delta \chi^\dagger}
\nonumber \\
&& ~~~~~~~~~~    + \left ( \partial A^a + \frac{1}{\alpha}
     Tr ( i \phi_v^\dagger T^a \phi - i \phi^\dagger T^a \phi_v )
     \right ) \frac{\delta \G^{(0)}}{\delta \bar \omega_a} \Bigg )  = 0\, .
\label{st1}
\end{eqnarray}
The linearized classical ST operator ${\cal S}_0$ is given by
\begin{eqnarray}
{\cal S}_0 & = & \int d^4x \, \Bigg (
\frac{\delta \G^{(0)}}{\delta A_\mu^{a*}}\frac{\delta}{\delta A^\mu_a}
+
\frac{\delta \G^{(0)}}{\delta A^\mu_a}\frac{\delta}{\delta A^{a*}_\mu}
+
\frac{\delta \G^{(0)}}{\delta \omega^{a*}}
\frac{\delta}{\delta \omega_a}
+
\frac{\delta \G^{(0)}}{\delta \omega_a}
\frac{\delta}{\delta \omega^{a*}}
\nonumber \\
&&
~~~~~~~~
+ Tr \frac{\delta \G^{(0)}}{\delta \chi^*} \frac{\delta}{\delta \chi}
+ Tr \frac{\delta \G^{(0)}}{\delta \chi} \frac{\delta}{\delta \chi^*}
+ Tr \frac{\delta \G^{(0)}}{\delta \chi^{*\dagger}} 
     \frac{\delta}{\delta \chi^\dagger}
+ Tr \frac{\delta \G^{(0)}}{\delta \chi^\dagger}
     \frac{\delta}{\delta \chi^{* \dagger}}
\nonumber \\
&& 
~~~~~~~~
+ \left ( \partial A^a + \frac{1}{\alpha}
     Tr ( i \phi_v^\dagger T^a \phi - i \phi^\dagger T^a \phi_v )
     \right ) \frac{\delta}{\delta \bar \omega_a} \Bigg ) \, .
\label{st2}
\end{eqnarray}
${\cal S}_0$ is nilpotent on the space of functionals obeying 
the ghost equation
\begin{eqnarray}
\frac{\delta \G^{(0)}}{\delta \bar \omega^a}
= -\alpha \partial^\mu \frac{\delta \G^{(0)}}{\delta A_{\mu a}^*}
- Tr \left [ i \phi_v^\dagger T^a \frac{\delta \G^{(0)}}{\delta \phi^*}
- i \frac{\delta \G^{(0)}}{\delta \phi^{\dagger *}} T^a \phi_v \right ] \, .
\label{st3}
\end{eqnarray}
%


\medskip

In the following subsection we will apply the embedding formalism
for the Higgs model 
to the example of the gauge group $SU(2)$ and
discuss the UV and IR properties of the propagators
after the shift  $\phi \rightarrow \phi + \phi_v$.

\subsection{The case of the $SU(2)$ gauge group}
\label{sect3}

We denote the Pauli matrices by $T^a$, $a=1,2,3$.
They obey the $SU(2)$ commutation relation
\begin{eqnarray}
[T^a,T^b]= 2 i \epsilon^{abc} T^c \, .
\label{nAb1}
\end{eqnarray}
By comparison with eq.(\ref{e1}) we get $f^{abc} = 2 \epsilon^{abc}$.
The field $\phi$ is a complex doublet which we parameterize as follows:
\begin{eqnarray}
\phi = \frac{1}{\sqrt{2}} \pmatrix{i \phi_1 + \phi_2 \cr \phi_0+v - i \phi_3}
\, .
\label{nAb2}
\end{eqnarray}
In a similar fashion we write for $b_\mu$
\begin{eqnarray}
b_\mu = \frac{1}{\sqrt{2}} \pmatrix{i b_{1\mu} + b_{2\mu} \cr
b_{0\mu} - i b_{3\mu} } \, .
\label{nAb3}
\end{eqnarray}
The covariant derivative acting on $\phi$ and on all
other fields in the same representation is
\begin{eqnarray}
D_\mu \phi (x) = \partial_\mu \phi(x) - i g A^a_\mu T^a \phi(x) \, ,
\label{nAb4}
\end{eqnarray}
where  we have restored the coupling constant $g$ in front
of $A^a_\mu$.
We choose a $R_\alpha$-gauge-fixing condition according to
\begin{eqnarray}
s \bar \omega^a = \partial A^a - \frac{gv}{\alpha} \phi^a \, .
\label{nAb5}
\end{eqnarray}
Now we can compute the quadratic part in the bosonic fields
in the classical action in eq.(\ref{em4}):
\begin{eqnarray}
\Sigma_2 & = & \int d^4x \, \Bigg ( 
+ \frac{1}{2} \sum_{a=1}^3 A_\mu^a
     ( (\square + (gv)^2) g^{\mu\nu} 
       - (1+\alpha) \partial^\mu \partial^\nu) A^a_\nu \nonumber \\
& & ~~~~~~~ 
+ \frac{1}{2} \partial_\mu \phi_0 \partial^\mu \phi_0 + \sum_{a=1}^3 \Big ( \frac{1}{2} \partial_\mu \phi_a \partial^\mu \phi_a + \frac{1}{2} \frac{(gv)^2}{\alpha} \phi_a^2 \Big ) \nonumber \\
         &   & ~~~~~~~
	       + \frac{1}{2} b^\mu_0 (\square^2 g_{\mu\nu} - (1 - \beta^2) \square  \partial_\mu \partial_\nu ) b^\nu_0 \nonumber \\
& & ~~~~~~~
	       + \frac{1}{2} \sum_{a=1}^3 b_a^\mu (\square^2 g_{\mu\nu} - (1 - \beta^2) \square \partial_\mu \partial_\nu ) b^\nu_a \nonumber \\
         &   & ~~~~~~~
               + \frac{gv}{2} \sum_{a=1}^3 \partial^{[\mu} b^{\nu]}_a \hat F_{\mu\nu}^a \Bigg ) \, ,
\label{nAb6}
\end{eqnarray}
where $\hat F^a_{\mu\nu}$ denotes the abelianization of the field strength
$F^a_{\mu\nu}$:
\begin{eqnarray}
\hat F^a_{\mu\nu} = \partial_\mu A^a_\nu - \partial_\nu A^a_\mu \, .
\label{nAb7}
\end{eqnarray}
Notice the appearance of the off-diagonal terms in the last line of
the eq.(\ref{nAb6}).

\medskip
We parameterize the antighosts $\psi_\mu$ and $\zeta$ and the ghosts 
$\xi_\mu, h$ as follows:
\begin{eqnarray}
& \psi_\mu = \frac{1}{\sqrt{2}} \pmatrix{-\psi_{1\mu} + i \psi_{2\mu} \cr
\psi_{3\mu} + i \psi_{0\mu}} \, , ~~~~ 
\zeta = \frac{1}{\sqrt{2}} \pmatrix{-\zeta_1 + i \zeta_2 \cr
\zeta_3 + i \zeta_0} \, , &
\nonumber \\
& \xi_\mu = \frac{1}{\sqrt{2}} \pmatrix{i \xi_{1\mu} + \xi_{2\mu} \cr
\xi_{0\mu} - i \xi_{3\mu}} \, , ~~~~
h = \frac{1}{\sqrt{2}} \pmatrix{ i h_1 + h_2 \cr h_0 - i h_3} \, .
 &
\label{nAbgh2}
\end{eqnarray}
Then the quadratic part in the ghost-antighost fields in eq.(\ref{em4})
is
\begin{eqnarray}
\Sigma_{gh,2} & = & \int d^4x \, \Bigg [ i
\Bigg ( -\psi_{0\mu} \Bigg ( (\square g^{\mu\rho} - \partial^\mu \partial^\rho) \xi_{0\rho} + \partial^\mu h_0 \Bigg ) + \beta \zeta_0 \partial \xi_0
\nonumber \\
&&
~~~~~~~~~ -\sum_{a=1}^3 \Bigg ( \psi_{a\mu} \Bigg ( 
(\square g^{\mu\rho} - \partial^\mu \partial^\rho) \xi_{a\rho} 
+ \partial^\mu h_a \Bigg ) - \beta \zeta_a \partial \xi_a \Bigg ) 
\Bigg )
\nonumber \\
&& 
~~~~~~~~~ +  \sum_{a=1}^3 \Bigg ( -\alpha \bar \omega_a \square \omega_a
+ (gv)^2 \bar \omega_a \omega_a + gv \bar \omega_a h_a 
\Bigg )
\Bigg ] \, .
\label{nAbgh3}
\end{eqnarray}

By inverting eq.(\ref{nAb6}) and eq.(\ref{nAbgh3}) we obtain the
UV $\delta$ and IR $\rho$ indices 
\cite{Zimmermann:1969jj}-\cite{Lowenstein:1975rg} of the fields,
which are given by
\begin{eqnarray}
& \delta(\phi_0) = \rho(\phi_0)=1 \, , ~~~~ 
\delta(b_{0\mu}) = \rho(b_{0\mu}) = 0 \, , &
\nonumber \\
& \delta(\phi_a) = 1 \, , ~ \rho(\phi_a) = 2 \, , ~~~~ a=1,2,3 \, ,&
\label{assign1}
\end{eqnarray}
and by
\begin{eqnarray}
&& \delta(\psi_{i\mu})=\rho(\psi_{i\mu})=1 \, ,
   ~~~ \delta(\xi_{i\mu})=\rho(\xi_{i\mu})=1 \,, \nonumber \\
&& \delta(h_i)=\rho(h_i)=2 \, , ~~~~~~
   \delta(\zeta_i)=\rho(\zeta_i)=2 \, , ~~~~~~~ i=0,1,2,3\, , \nonumber \\
&& \delta(\bar \omega_a) = \delta(\omega_a)=1 \, , ~~~~~
   \rho(\bar \omega_a)= \rho(\omega_a) = 2 \, , ~~~~~ a = 1,2,3 \,.
\label{assign2}
\end{eqnarray} 
In the sector $A^a_\mu-b^c_{\nu}$ we find the following propagators
\footnote{ for notational convenience we omit to write the $i\epsilon$
factor.}
\begin{eqnarray}
\Delta_{b^c_{\mu}(-p)b^d_{\nu}(p)} = \delta^{cd}
\frac{-i}{(p^2)^2 F(p^2)} \left ( g_{\mu\nu} + \frac{F(p^2) - \beta^2}{\beta^2}
\frac{p_\mu p_\nu}{p^2} \right ) \, , 
\label{assign4}
\end{eqnarray}
where we have set
\begin{eqnarray}
F(p^2) = 1 - \frac{g^2v^2}{-p^2 + (gv)^2} \, ,
\label{ab8}
\end{eqnarray}
and
\begin{eqnarray}
\Delta_{A^a_\mu(-p) b^d_{\nu}(p)} = \delta^{ad} 
\frac{igv}{(p^2)^3}\left ( -p^2 g_{\mu\nu} + p_\mu p_\nu \right ) \, , 
\label{assign5}
\end{eqnarray}
\begin{eqnarray}
\Delta_{A^a_\mu(-p) A^b_\nu(p)} = \Delta_{A^{'a}_\mu(-p) A^{'b}_\nu(p)}
+ \delta^{ab} \frac{i (gv)^2}{( p^2 - (gv)^2)^2}\frac{1}{p^2 F(p^2)} ( -p^2 g_{\mu\nu}
+ p_\mu p_\nu) 
\label{assign6}
\end{eqnarray}
with
\begin{eqnarray}
\Delta_{A^{'a}_\mu(-p) A^{'b}_\nu(p)} = \delta^{ab} 
\frac{i}{p^2 - (gv)^2} \left (
g_{\mu\nu} - \frac{(1+\alpha)}{\alpha p^2 + (gv)^2} p_\mu p_\nu \right ) \, .
\label{assign7}
\end{eqnarray}
Since $F(p^2)$ tends to $1$ for large $p$ the UV degrees are 
\begin{eqnarray}
\delta(A^a_\mu)=1 \, , ~~~~ \delta(b^a_{\nu})=0 \,,
\label{assign8}
\end{eqnarray}
compatible with the power-counting renormalizability criteria.
In the infrared regime the behavior of the propagators in
eqs.(\ref{assign4})-(\ref{assign6}) yields
the assignments
\begin{eqnarray}
\rho(A^a_\mu)=1 \, , ~~~~ \rho(b^b_{\nu})=-1 \, ,
\label{assign9}
\end{eqnarray}
which then lead to a bad IR power-counting (the classical action
contains vertices with IR degree $<4$). This destroys
the renormalizability  of the model by power-counting.

\medskip

Let us comment on the results obtained so far. We have discussed
the embedding of the Higgs model into a larger
theory which is invariant under the BRST differential $s$
in eq.(\ref{e12}).
No quartic potential for the field $\phi$ is any longer allowed,
since it is forbidden by the shift
symmetry $s_1$.
The analysis of the propagators and the interaction vertices
shows that the UV conditions for power-counting renormalizability
are fulfilled, while the IR ones are not. 
This is due to a bad IR asymptotic behaviour of the propagators 
in the sector spanned by the fields $A_\mu^a$ and $b_\nu^b$.
As a consequence, the Higgs model embedding is 
not power-counting renormalizable.

\medskip
We conclude that the embedding procedure in the presence of the
Higgs mechanism fails to produce the sought-for 
power-counting renormalizable
embedding of massive Yang-Mills theory.

Nevertheless,
despite its failure the model just analyzed
provides a clue
towards the correct embedding of massive Yang-Mills theory.
In this connection we would like to comment on the shift symmetry fixing terms
given by the $s_1$-variations of $\psi_\mu,\psi_\mu^\dagger$ and 
$\zeta,\zeta^\dagger$ 
in eq.(\ref{e13}).

According to eqs.(\ref{e11})
and (\ref{e11bis}) $s$ squares to zero only modulo
the equations of motion of $\psi_\mu,\psi^\dagger_\mu$ and
$\zeta,\zeta^\dagger$. Both eqs.(\ref{e11}) and (\ref{e11bis})
can be dealt with by the BV formalism. 
Nevertheless there exists an important difference between eq.(\ref{e11})
and eq.(\ref{e11bis}): the first one can be implemented
in an off-shell formulation by introducing suitable
Nakanishi-Lautrup multiplier fields for $\psi_\mu,\psi_\mu^\dagger$,
the second one does not admit such an off-shell formulation.

\medskip
This suggests that there should exist another embedding
procedure where the fields $\zeta,\zeta^\dagger$ 
(and consequently $\phi,\phi^\dagger$ and
$h,h^\dagger$) do not appear.
Such a minimal embedding model can actually be constructed.
It turns out that it naturally solves the IR problems affecting
the Higgs model embedding. 
Moreover, we will prove that 
 it gives rise to a true power-counting renormalizable model
for massive non-Abelian gauge bosons without spontaneous
gauge symmetry breaking.
As a consequence of the embedding BRST symmetry, the physical observables 
are given by gauge-invariant quantities constructed only from the field
strength and its covariant derivatives.

The analysis of the minimal embedding formalism will be
carried out in the next section.

\section{Minimal embedding}\label{minim}

In the minimal embedding formalism we drop the fields
$\phi,\phi^\dagger$ and their ghost partners $h,h^\dagger$ together
with the antighost fields $\zeta,\zeta^\dagger$.

The fixing action $S$ is now given by
\begin{eqnarray}
S & = & Tr \int d^4x \, K_\mu^\dagger K^\mu \nonumber \\
  &   & + Tr \int d^4x ( \psi^\dagger_\mu s K^\mu + 
                         (s K^\mu)^\dagger \psi_\mu ) \, , 
\label{min1}
\end{eqnarray}
where 
\begin{eqnarray}
K_\mu \equiv D_\mu (Db) + \lambda D^2 b_\mu + \gamma \frac{m^2}{2} b_\mu
                              +D_\mu \underline{M} \, .
\label{min2}
\end{eqnarray}
In the above equation we have set
\begin{eqnarray}
\underline{M} = \frac{1}{\sqrt{2}} \underline{m} \, .
\label{min3}
\end{eqnarray}
with $\underline{M},\underline{m}$ 
in the same representation as $b_\mu$.
For instance, in the case of the $SU(2)$ group
we get
\begin{eqnarray}
\underline{M} = \frac{1}{\sqrt{2}} \pmatrix{0 \cr m} \, .
\label{min3_bis}
\end{eqnarray}
$m$ is a parameter with the dimension of a mass.

$\lambda,\gamma$ are real parameters. We will comment
later on on their physical significance.
The BRST transformations of $b^\mu$ and their ghost partners
$\xi^\mu$ are the same as in eqs.(\ref{e6}) and (\ref{e7}),
while
\begin{eqnarray}
s \psi_\mu = K_\mu \, , ~~~~ s \psi_\mu^\dagger = - K_\mu^\dagger \, .
\label{min4}
\end{eqnarray}
The explicit form of $s K_\mu$ is
\begin{eqnarray}
s K_\mu & = & i \omega_a T^a 
              (D_\mu (Db) + \lambda D^2 b_\mu + \gamma \frac{m^2}{2} b_\mu ) \nonumber \\
        &   & + D_\mu (D\xi) + \lambda D^2 \xi_\mu + \gamma \frac{m^2}{2} \xi_\mu \nonumber \\
        &   & -i (D_\mu \omega)_a T^a \underline{M} \, .
\label{min5}
\end{eqnarray}
We notice that gauge invariance is explicitly broken
by the last term in eq.(\ref{min5}). In this model there is no
spontaneous gauge symmetry breaking.

$S$ in eq.(\ref{min1}) is $s$-invariant.
$s$ squares to zero modulo the equations of motion
of $\psi^\mu,\psi^{\mu \dagger}$:
\begin{eqnarray}
s^2 \psi_\mu = \frac{\delta S}{\delta \psi^{\mu\dagger}} \, , ~~~~
s^2 \psi^{\dagger}_\mu = \frac{\delta S}{\delta \psi^\mu} \, .
\label{min6}
\end{eqnarray}
Due to eq.(\ref{min6}) an off-shell formulation of the model
exists, based on the introduction of the Nakanishi-Lautrup multiplier
fields which form BRST doublets together with $\psi_\mu,\psi_\mu^\dagger$.
Hence $S$ turns out to be a true fixing functional for the 
symmetry $s$. 

In what follows we choose to work in the on-shell formalism and we shall
use the BV method in order to derive the relevant ST identities
of the model.

The embedding action reads
\begin{eqnarray}
S_{\rm embed} = -\frac{1}{4g^2} \int d^4x \, F_{\mu\nu}^a F^{\mu\nu a} + S \,.
\label{min7}
\end{eqnarray}
This action is $s$-invariant. Notice that we have not yet chosen 
the gauge-fixing function for the vector fields $A_\mu^a$.

\subsection{Analysis of the UV and IR properties of the model}
\label{subMin_IR}

In this subsection we analyze the UV and IR properties
of the minimal embedding model by deriving the UV and IR
degrees of the fields and by checking that
the interaction vertices obey the conditions
for power-counting renormalizability 
\cite{Zimmermann:1969jj,Lowenstein:1975ps,Lowenstein:ug,Lowenstein:1975rg}.

For the sake of definiteness we specialize from now on to the
case of the gauge group $SU(2)$. Nevertheless the analysis
can be straightforwardly extended to other gauge groups of interest,
including those involving Abelian factors like $SU(2) \times U(1)$.

We use the same conventions about normalizations and
covariant derivatives as in Sect.~\ref{sect3}.

\subsubsection{The bosonic sector}

The quadratic part of $S_{\rm embed}$ in eq.(\ref{min7})
 is given in the bosonic sector by
\begin{eqnarray}
S_{\rm embed,II} & = & \int d^4x \, \Big ( -\frac{1}{4} \sum_{a=1}^3 
\hat F_{\mu\nu}^a\hat F^{\mu\nu \, a} \nonumber \\
& & + \frac{1}{2} \big ( \partial_\mu (\partial b_0) + \lambda \square b_{0\mu} + \frac{\gamma m^2}{2} b_{0\mu} \big )^2
\nonumber \\
& & + \sum_{a=1}^3 
\frac{1}{2} \big ( \partial_\mu (\partial b_a) + \lambda \square b_{a\mu} + \frac{\gamma m^2}{2} b_{a\mu} 
                        -gm A_{a\mu} \big )^2 \Big ) \, ,
\label{min_eq20}
\end{eqnarray}
where $\hat F_{\mu\nu}^a$ denotes the Abelianization of the 
field strength $F_{\mu\nu}^a$:
\begin{eqnarray}
\hat F_{\mu\nu}^a = \partial_\mu A_\nu^a - \partial_\nu A_\mu^a \, .
\label{min_ab}
\end{eqnarray}
We see from eq.(\ref{min_eq20}) that for non-exceptional values of 
the parameters $\lambda$ and $\gamma$ the UV and IR degrees of $b_{0\mu}$ are
\begin{eqnarray}
\delta(b_{0\mu})=0 \, , ~~~~ \rho(b_{0\mu})=2 \, .
\label{min_eq24}
\end{eqnarray}

\medskip

The quadratic part of $S_{\rm embed}$ in the sector $A_{\mu a}-b_{\nu b}$ is
\begin{eqnarray}
S_{\rm embed,II,A-b}  & = & \int d^4x \, \Bigg [ \sum_{a=1}^3 
\frac{1}{2} b_{a\mu} 
\Bigg [ \Big ( \lambda^2 \square^2 + \lambda \gamma m^2 \square + \left ( \frac{\gamma m^2}{2} \right )^2 \Big ) g^{\mu \nu}
\nonumber \\ 
&&  ~~~~~~~~~~~ + ( (1 + 2\lambda) \square + \gamma m^2 ) \partial^\mu \partial^\nu \Bigg ] b_{a\nu} \nonumber \\
&&  ~~~~~~~~ + \sum_{a=1}^3 
\frac{1}{2} A_\mu^a \Big ( (\square + g^2m^2) g^{\mu\nu} 
                - \partial^\mu \partial^\nu \Big ) A^a_\nu 
\nonumber \\
&&  ~~~~~~~~ + \sum_{a=1}^3 \Big ( gm \,  \partial A^a \partial b^a 
             - g m A^{\mu a} \Big ( \lambda \square + \frac{\gamma m^2}{2} \Big )  
               b^{a}_\mu \Big ) \Bigg ] \, . \nonumber \\
\label{min_eq25}
\end{eqnarray}
The mixed term $\partial A_a \partial b_a$ can be removed by making use
of the following $R_\alpha$~-gauge-fixing:
\begin{eqnarray}
\frac{\alpha}{2} \left ( \partial A_a - \frac{gm}{\alpha} \partial b_a \right )^2
\label{min_eq26}
\end{eqnarray}
Then eq.(\ref{min_eq25}) becomes
\begin{eqnarray}
S_{\rm embed,II,A-b}  & = & \int d^4x \, \Bigg [ \sum_{a=1}^3 
\frac{1}{2} b_{a\mu} 
\Bigg [ \Big ( \lambda^2 \square^2 + \lambda \gamma m^2 \square + \left ( \frac{\gamma m^2}{2} \right )^2 \Big ) g^{\mu \nu}
\nonumber \\ 
&&  ~~~~~~~~~~~ + ( (1 + 2\lambda) \square + \gamma m^2 - \frac{(gm)^2}{\alpha} ) \partial^\mu \partial^\nu \Bigg ] b_{a\nu} \nonumber \\
&&  ~~~~~~~~ + \frac{1}{2} \sum_{a=1}^3 
A_{\mu a} \Big ( (\square + g^2m^2) g^{\mu\nu} 
                - (1 + \alpha) \partial^\mu \partial^\nu \Big ) A_{a \nu}
\nonumber \\
&&  ~~~~~~~~ 
- \sum_{a=1}^3 g m A_{a\mu} \Big ( \lambda \square  + \frac{\gamma m^2}{2} \Big ) b_a^\mu \Bigg ] \, .
\label{min_eq26_bis}
\end{eqnarray}
By inverting the two-point function matrix in
eq.(\ref{min_eq26_bis})  we obtain the following
assignments of the UV and IR indices:
\begin{eqnarray}
&& \delta(A_{a \mu}) = 1 \, , ~~~~ \rho(A_{a\mu})=1 \, , 
\nonumber \\
&& \delta(b_{a \mu}) = 0 \, , ~~~~ \rho(b_{a\mu})=1 \, .
\label{min_eq32}
\end{eqnarray}

\subsubsection{The ghost sector}

The gauge-fixing choice in eq.(\ref{min_eq26}) implies that
\begin{eqnarray}
s \bar \omega^a = \partial A^a - \frac{gm}{\alpha} \partial b^a \, .
\label{min_eq33}
\end{eqnarray}
Therefore the $\bar \omega^a$-dependent part of the classical action
reads
\begin{eqnarray}
- \int d^4x \, \alpha \bar \omega^a \Big (
\partial_\mu (D^\mu \omega)^a - \frac{gm}{\alpha} \partial \xi^a
-\frac{g^2m}{\alpha} ( \partial^\mu(\omega^a b_{0\mu}) 
- \epsilon^{abc} \partial^\mu (\omega_b b_{c \mu}) ) \Big ) \, .
\label{min_eq34}
\end{eqnarray}

The quadratic part in the ghost sector is then given by
\begin{eqnarray}
\Sigma_{{\rm II,ghost}} & = &
\int d^4x \, \Bigg ( - i \sum_{a=1}^3 
            \psi_{a\mu} \Big ( \partial^\mu (\partial \xi_a)
            +\lambda \square \xi_a^\mu + \frac{1}{2} \gamma m^2 \xi_a^\mu
            - gm \partial^\mu \omega_a \Big ) \nonumber \\
& & - i \psi_{0 \mu} \Big ( \partial^\mu (\partial \xi_0)
            +\lambda \square \xi_0^\mu + \frac{1}{2} \gamma m^2 \xi_0^\mu 
            \Big ) \nonumber \\
& & - \alpha \bar \omega_a \Big ( \square \omega_a - \frac{gm}{\alpha} \partial \xi_a \Big ) 
\Bigg ) \, .     
\label{min_gh5}
\end{eqnarray}

The inversion of the ghost two-point function matrix in eq.(\ref{min_gh5})
yields the following assignments of 
the UV and IR indices:
\begin{eqnarray}
&& 
\!\!\!\!\!\!\!\!\!\!
\delta(\psi_{0\mu})=\delta(\psi_{a\mu})=\delta(\xi_{0\mu})
=\delta(\xi_{a\mu}) = 1 \, , ~~~~
\delta(\bar \omega^a)=\delta(\omega^a) = 1 \, , \nonumber \\
&& 
\!\!\!\!\!\!\!\!\!\!
\rho(\psi_{0\mu})=\rho(\psi_{a\mu})=\rho(\xi_{0\mu})
=\rho(\xi_{a\mu}) = 2 \, , ~~~~
\rho(\bar \omega^a)=\rho(\omega^a) = 1 \, .
\label{min_gh6}
\end{eqnarray}

A table summarizing the UV and IR degrees of all the fields
of the model is reported in Appendix~\ref{appB}.

\medskip
By using the IR and UV assignments in eq.(\ref{min_eq24}), 
(\ref{min_eq32}) and (\ref{min_gh6}) one can
verify that all the interaction vertices in $S_{\rm embed}$ in eq.(\ref{min7})
have UV degree $\leq 4$ and IR degree $\geq 4$. Hence
they satisfy the power-counting conditions
\cite{Zimmermann:1969jj,Lowenstein:1975ps,Lowenstein:ug,Lowenstein:1975rg}
which are sufficient 
to ensure the power-counting renormalizability of the model.
The minimal embedding procedure thus naturally solves the IR difficulties
discussed in Sect.~\ref{sect3} for the Higgs model embedding.

\subsection{Classical Slavnov-Taylor identities}

Since 
in the on-shell formalism
the BRST differential $s$ squares to zero only modulo
the equations of motions of $\psi_\mu^\dagger,\psi_\mu$,
we use the BV method to obtain the relevant ST identities
of the model.

The classical action $\G^{(0)}$ fulfilling these ST identities
includes a term quadratic in the antifields 
$\psi_\mu^{\dagger *},\psi_\mu^*$.
$\G^{(0)}$ is given by
\begin{eqnarray}
\G^{(0)} & = & \int d^4x \, \Bigg ( -\frac{1}{4} F_{\mu\nu}^a F^{\mu\nu a}
+ \frac{\alpha}{2} \Big ( \partial A^a - \frac{gm}{\alpha} \partial b^a)^2
\nonumber\\
& & + Tr \Big ( K_\mu^\dagger K^\mu + \psi_\mu^\dagger s K^\mu
               + (s K_\mu)^\dagger  \psi^\mu \Big ) \nonumber \\
& & -\alpha \bar \omega^a \Big ( \partial^\mu (D_\mu \omega)^a 
- \frac{gm}{\alpha} \partial \xi^a - \frac{g^2 m }{\alpha}
\partial^\mu ( \omega^a b_{0\mu} - \epsilon^{abc} \omega^b b^c_\mu ) \Big )
\nonumber \\
& & + A_\mu^{a*} (D^\mu \omega)^a - \omega^{a*} \frac{1}{2} g f^{abc} 
\omega^b \omega^c \nonumber \\
& & + Tr \Big ( b_\mu^* \, s b^\mu + b_\mu^{\dagger *} \, s b^{\dagger \mu}
               +\xi_\mu^* \, s \xi^\mu + \xi_\mu^{\dagger *} \, s \xi^{\dagger \mu} \Big ) \nonumber \\
& & + Tr \Big ( \psi_\mu^* K^\mu - \psi_\mu^{\dagger *} K^{\mu \dagger}
- \psi_\mu^{\dagger *} \psi^{\mu *} \Big ) \Bigg ) \, .
\label{eq1_1}
\end{eqnarray}

No differential operators enter in the last term in eq.(\ref{eq1_1}),
unlike in the case of the last monomial in eq.(\ref{em4}).

\medskip
$\G^{(0)}$ fulfills the following ST identities
\begin{eqnarray}
{\cal S}(\G^{(0)}) & = & \int d^4 x \, \Bigg (
\frac{\delta \G^{(0)}}{\delta A_\mu^{a*}}
\frac{\delta \G^{(0)}}{\delta A_\mu^{a}}
+
\frac{\delta \G^{(0)}}{\delta \omega^{a*}}
\frac{\delta \G^{(0)}}{\delta \omega^{a}}
\nonumber \\
 & & ~~~~~~~ + Tr \frac{\delta \G^{(0)}}{\delta \chi^*}
     \frac{\delta \G^{(0)}}{\delta \chi} \nonumber \\
 & & ~~~~~~~ + \Big ( \partial A^a - \frac{gm}{\alpha} \partial b^a \Big )
  \frac{\delta \G^{(0)}}{\delta \bar \omega^a} \Bigg ) = 0 \, .
\label{eq1_2}
\end{eqnarray}
In the above equation $\chi$ stands for 
$\chi = \{ b_\mu, b_\mu^\dagger, \xi_\mu, \xi^\dagger_\mu, \psi_\mu,
\psi_\mu^\dagger \}$ and
$\chi^*$ for
$\chi^* = \{ b_\mu^*, b_\mu^{\dagger *}, \xi^*_\mu, \xi^{\dagger *}_\mu, 
\psi_\mu^*, \psi_\mu^{\dagger *} \}$.
Notice that we do not introduce an antifield for $\bar \omega^a$,
since its BRST transformation is linear in the quantum fields.

\medskip
The classical linearized ST operator ${\cal S}_0$ is given by
\begin{eqnarray}
\!\!\!\!
{\cal S}_0 & = & \int d^4x \, \Bigg (
\frac{\delta \G^{(0)}}{\delta A_\mu^{a*}}
\frac{\delta}{\delta A_\mu^{a}}
+
\frac{\delta \G^{(0)}}{\delta A_\mu^{a}}
\frac{\delta}{\delta A_\mu^{a*}}
+
\frac{\delta \G^{(0)}}{\delta \omega^{a*}}
\frac{\delta}{\delta \omega^{a}}
+
\frac{\delta \G^{(0)}}{\delta \omega^{a}}
\frac{\delta }{\delta \omega^{a*}}
\nonumber \\
& & 
 ~~~~~~~+
Tr \frac{\delta \G^{(0)}}{\delta \chi^*}
     \frac{\delta}{\delta \chi}
+
Tr \frac{\delta \G^{(0)} }{\delta \chi}\frac{\delta}{\delta \chi^*}
+ \Big ( \partial A^a - \frac{gm}{\alpha} \partial b^a \Big )
  \frac{\delta}{\delta \bar \omega^a} \Bigg )  \, .
\label{eq1_3}
\end{eqnarray}
${\cal S}_0$ is nilpotent on the space of functionals satisfying the ghost
equation
\begin{eqnarray}
\frac{\delta \G^{(0)}}{\delta \bar \omega_a} 
= - \alpha \partial^\mu \frac{\delta \G^{(0)}}{\delta A_\mu^{a*}}
- \frac{gm}{\alpha} \partial^\mu \frac{\delta \G^{(0)}}{\delta b_\mu^{a*}} \, .
\label{eq1_4}
\end{eqnarray}

\subsection{Analysis of the cohomology of ${\cal S}_0$}\label{cohom}

All the fields and antifields of the model with the exception
of $A_\mu^a,\omega^a$ and their antifields and $\bar \omega^a$
pair into ${\cal S}_0$-doublets. Hence the cohomology
$H({\cal S}_0)$ in the space of functionals fulfilling the ghost
equation
\begin{eqnarray}
\frac{\delta X}{\delta \bar \omega^a} 
= - \alpha \partial^\mu \frac{\delta X}{\delta A_\mu^{a*}}
- \frac{gm}{\alpha} \partial^\mu \frac{\delta X}{\delta b_\mu^{a*}} 
\label{eq1_5}
\end{eqnarray}
is isomorphic to the cohomology of the restriction ${\cal S}_0'$
of ${\cal S}_0$ to the space of functionals spanned by
$A_\mu^a,\omega^a$ and their antifields and by $\bar \omega^a$,
which obey eq.(\ref{eq1_5}) \cite{Barnich:2000zw,Piguet:er,Quadri:2002nh}.

The cohomology of such a restriction is known 
to be isomorphic \cite{Barnich:2000zw,Piguet:er} to the cohomology $H(s_0)$ 
of the pure Yang-Mills differential $s_0$: 
\begin{eqnarray}
s_0 A_\mu^a = (D_\mu \omega)^a \, , ~~~~
s_0 \omega^a = -\frac{1}{2} g f^{abc} \omega^b \omega^c \, .
\label{pureym01}
\end{eqnarray}
In the FP-neutral sector (physical observables) 
$H(s_0)$ is given by
the gauge-invariant quantities constructed only from the
field strength $F_{\mu\nu}^a$ and its covariant derivatives.
%
These are the physical observables of the minimal embedding model.

\medskip 
In the sector with FP-charge $+1$ (anomalies) the cohomology
of ${\cal S}'_0$ in the relevant functional space fulfilling
eq.(\ref{eq1_5}) is empty \cite{Barnich:2000zw,Piguet:er}.
Hence there are no candidate anomalies for the quantum extension
of the ST identities in eq.(\ref{eq1_2}).

Together with power-counting renormalizability, guaranteeing
the validity of the Quantum Action Principle \cite{Piguet:er}, this implies
that the quantum ST identities
\begin{eqnarray}
{\cal S}(\GG) = 0 
\label{eq1_6}
\end{eqnarray}
can be restored order by order in the loop expansion by a 
suitable choice of finite counterterms, yielding the symmetric
quantum effective action
\begin{eqnarray}
\GG = \sum_{j=0}^\infty \GG^{(j)}
\label{eq1_7}
\end{eqnarray}
which fulfills eq.(\ref{eq1_6}).
In eq.(\ref{eq1_7}) $\GG^{(j)}$ stands for the coefficient of order
$j$ in the loop expansion of $\GG$. $\GG^{(0)}$ coincides
with $\G^{(0)}$ in eq.(\ref{eq1_2}).

\subsection{The spectrum of the classical theory}
\label{spectrum}

As a consequence of the cohomological analysis developed
in the previous subsection, we know that
the physical observables of the minimal embedding model
are given by gauge-invariant quantities constructed only
from the field strength $F_{\mu\nu}^a$ and its covariant derivatives.

We wish to complete the analysis of the physical content of the theory
by computing the physical spectrum of the classical action in eq.(\ref{eq1_1}).
This will also clarify the meaning of the parameters
$\lambda$ and $\gamma$ in eq.(\ref{min2}).
For this purpose
the study of the relevant asymptotic states is needed.

\medskip
We first analyze the sector spanned by $A_\mu^a$ and $b_\nu^c$.
From now on we choose to work in the Feynman gauge $\alpha=-1$.
The quadratic part in $A_\mu^a,b_\nu^c$ in eq.(\ref{min_eq26_bis})
is given in the momentum space by
\begin{eqnarray}
\!\!\!\!\!\!\!\!
S_{\rm embed,II,A-b}  & = & \int d^4p \, \Bigg [ \sum_{a=1}^3
\frac{1}{2} b_{a\mu}(-p)
\Bigg [ \Big ( \lambda p^2 - \frac{\gamma m^2}{2} \Big )^2 g^{\mu\nu}
\nonumber \\
&& + \Big ((1+2\lambda)p^2 - \gamma m^2  -(gm)^2 \Big ) p^\mu p^\nu 
\Bigg ] b_{a\nu}(p) \nonumber \\
&& + \sum_{a=1}^3 \frac{1}{2} A_{a\mu}(-p)  (-p^2 + g^2m^2)  A_{a}^\mu(p) \nonumber \\
&& + \sum_{a=1}^3
A_{a\mu}(-p) gm \Big (  \lambda p^2 - \frac{\gamma m^2}{2} \Big ) b_a^{\mu}(p) 
\Bigg ] \, .
\label{min_eq30}
\end{eqnarray}
The asymptotic states are the elements of the kernel of
the two-point function matrix 
$\underline{\Sigma}=\Sigma_{(\mu,a,L; ~ \mu',a',L')}$ computed
from eq.(\ref{min_eq30}):
\begin{eqnarray}
\Sigma_{(\mu,a,L; ~ \mu',a',L')} = 
\frac{\delta^2 S_{\rm embed,II,A-b}}
     {\delta \varphi_\mu^{a,L}(-p) \delta \varphi_{\mu'}^{a',L'}(p)} 
\label{app1}
\end{eqnarray}
where we have denoted by $\varphi_\mu^{a,L}$ the column vector
with components
$$\varphi_\mu^{a,1}=A_\mu^a \, , ~~~~
  \varphi_\mu^{a,2}=b_\mu^a \, .$$
$\underline{\Sigma}$ is diagonal in the $a a'$-space.

\medskip

The values of the masses in the spectrum are given by
the zeroes of $\Sigma= {\rm det}~ \underline{\Sigma}$.
For arbitrary values of $\lambda$ and $\gamma$
$\Sigma$ vanishes at
\begin{eqnarray}
p^2=0\, , ~~~~ p^2=\frac{\gamma m^2}{2\lambda} \, , ~~~~
p^2= \frac{(2g^2 + \gamma)m^2}{2(1 + \lambda)}.
\label{poles}
\end{eqnarray}

The requirement that the spectrum includes the point
$p^2 = g^2 m^2$ imposes a constraint on $\gamma$ and $\lambda$:
\begin{eqnarray}
\gamma = 2 g^2 \lambda \, .
\label{app2}
\end{eqnarray}
With the choice in eq.(\ref{app2})
it turns out that $p^2=0$ is a zero of $\Sigma$ of order four,
while $p^2= g^2m^2$ is a zero of order eight.
It can be seen that if we impose the degeneracy of the second and
third pole in eq.(\ref{poles}) their common value is uniquely
given by $p^2 = g^2m^2$.

\medskip

\medskip
If eq.(\ref{app2}) holds,
diagonalization of the quadratic form in eq.(\ref{min_eq30})
can be achieved by a local field redefinition.
For $\gamma=2 g^2 \lambda$ 
eq.(\ref{min_eq30}) becomes
\begin{eqnarray}
S_{\rm embed,II,A-b}  & = & 
\int d^4x \, \Bigg [ \sum_{a=1}^3 \frac{1}{2} b_\mu^a
\Big ( \lambda^2 (\square + g^2m^2)^2 g^{\mu\nu} \nonumber \\
& & ~~~~~~~~~ + 
(1 + 2\lambda) (\square + g^2m^2) \partial^\mu \partial^\nu \Big ) b_\nu^a
\nonumber \\
& & ~~~~~~~~~ + \sum_{a=1}^3 \, \frac{1}{2} A_\mu^a (\square + g^2m^2) A^{\mu a}
\nonumber \\
& &  ~~~~~~~~~
- \lambda g m \sum_{a=1}^3 A_\mu^a (\square + g^2 m^2) b^{\mu a} \Bigg ] \, .
\label{min_eq30_1}
\end{eqnarray}

The diagonalization is obtained by setting
\begin{eqnarray}
A_{a\mu} = A'_{a\mu} + \lambda \, gm \,  b_{a\mu} \, .
\label{eq1_10}
\end{eqnarray}
This yields finally
\begin{eqnarray}
S_{\rm embed,II,A-b}  & = & 
\int d^4x \, \Bigg [ \sum_{a=1}^3 \frac{1}{2} 
b_{\mu a} \Big ( \lambda^2 (\square + g^2m^2) \square g^{\mu\nu} \nonumber \\
& & ~~~~~~~~~ 
+ (1 + 2\lambda) (\square + g^2 m^2) \partial^\mu \partial^\nu \Big )
b_{\nu a} \nonumber \\
& & ~~~~~~~~~ + \frac{1}{2} \sum_{a=1}^3 A'_{\mu a} (\square + g^2 m^2) A^{' \mu}_a \Bigg ] \, .
\label{eq1_11}
\end{eqnarray}
Without the condition in eq.(\ref{app2}) it turns out that
it is not possible to diagonalize eq.(\ref{min_eq30}) by a local
field redefinition. Therefore the physical interpretation of the
corresponding model  is much less transparent and will not be pursued
here.

The UV and IR assignments in 
eqs.(\ref{min_eq24}), (\ref{min_eq32}) and (\ref{min_gh6})
are also valid
for the special choice $\gamma= 2 g^2 \lambda$, $\alpha=-1$.

A comment is in order here. The off-diagonal terms in the second line
of eq.(\ref{eq1_11}) disappear if one further chooses
\begin{eqnarray}
\lambda= - \frac{1}{2} \, .
\label{diag}
\end{eqnarray}
However
 the condition of the absence of poles at
negative values in the propagator of $b^0_\mu$ yields
an exclusion region for $\lambda$ given by $-1 < \lambda < 0$,
thus forbidding the choice in eq.(\ref{diag}).

\medskip
In the limit $\lambda \rightarrow 0$ we recover under
the formal identification $\phi = Db$ the classical
Higgs model in the absence of the quartic potential,
as it can be seen from eq.(\ref{min1}) and eq.(\ref{min2}).
In this limit the scalar $\partial b^0$, corresponding to the
physical Higgs field, is massless.
On the other hand in the limit $\lambda \rightarrow -1$ 
the scalar $\partial b^0$ becomes infinitely massive.

\medskip
Since $S_{\rm embed,II,A-b}$ in eq.(\ref{eq1_11}) contains
the dipole field $b_\mu^a$, a consistent analysis of the asymptotic
states requires the use of a second order formalism
relying on the introduction  of suitable auxiliary fields
\cite{Narnhofer:sw}.

Two different formulations are possible. 
In the first one we rewrite eq.(\ref{eq1_11}) as follows
(for notational convenience we suppress the sum over the index $a$):
\begin{eqnarray}
S_1 & = & \int d^4x \, \left ( \frac{1}{2} A_a^{'\mu} (\square + g^2m^2) A^{'}_{a \mu}
       + h_{a\mu} [ C_a^\mu - (\square + g^2m^2) b_a^\mu ] \right .
       \nonumber \\
  &   & ~~~~~~~~~~~ \left . 
       + \frac{1}{2} C_{a\mu} [ \lambda^2 \square b_a^\mu + 
         (1+2 \lambda) \partial^\mu (\partial b_a) ] \right ) \, .
\label{note1}
\end{eqnarray}
The equations of motion for $h_{a\mu}, C_{a\mu}$ and  $b_{a\mu}$ are
\begin{eqnarray}
&& \frac{\delta S_1}{\delta h_{a\mu}} = C_a^\mu - (\square + g^2m^2) b_a^\mu = 0 \, ,
\nonumber \\
&& \frac{\delta S_1}{\delta C_{a\mu}} = h_a^\mu 
   + \frac{1}{2}  [ \lambda^2 \square b_a^\mu + 
     (1+2 \lambda) \partial^\mu (\partial b_a) ] = 0 \, , \nonumber \\
&& \frac{\delta S_1}{\delta b_{a\mu}} = - (\square + g^2 m^2) h_a^\mu 
   + \frac{1}{2} \lambda^2 \square C_a^\mu 
   + \frac{1}{2} (1 + 2\lambda) \partial^\mu (\partial C_a) = 0 \, .
\label{note2}
\end{eqnarray}
By going on-shell in eq.(\ref{note1})  with $h_{a\mu}$ $S_1$
reduces to $S_{\rm embed,II,A-b}$ in eq.(\ref{eq1_11}).

\medskip
The second possibility is to consider 
\begin{eqnarray}
S_2 & = & \int d^4x \, \Big ( 
\frac{1}{2} A^{'\mu}_a (\square + (gm)^2) A'_{a \mu}
+ \h_{a\mu} (\C^\mu_a - \lambda^2 \square b^{\mu}_a -
(1 + 2\lambda) \partial^\mu (\partial b_a) ) \nonumber \\
& & ~~~~~~~ + \frac{1}{2} \C_{a\mu} (\square + (gm)^2) b_a^\mu \Big ) \, .
\label{note3}
\end{eqnarray}
The equations of motion are 
\begin{eqnarray}
\frac{\delta S_2}{\delta \h_{a\mu}} & = & \C_a^\mu -
      \lambda^2 \square b_a^\mu - (1+2\lambda) \partial^\mu(\partial b_a) = 0\, , 
\nonumber \\
\frac{\delta S_2}{\delta \C_{a\mu}} & = & \h_a^\mu 
      + \frac{1}{2} (\square + (gm)^2)b_a^\mu = 0 \, , 
\nonumber \\
\frac{\delta S_2}{\delta b_{a\mu}} & = & 
- (\lambda^2 \square \h_a^\mu + (1+2\lambda) \partial^\mu(\partial \h_a))
+ \frac{1}{2} (\square + (gm)^2) \C_a^\mu = 0 \, .
\label{note4}
\end{eqnarray}
By going on-shell in eq.(\ref{note3}) with $\h_{a\mu}$ 
$S_2$ reduces to $S_{\rm embed,II,A-b}$ in eq.(\ref{eq1_11}).

We notice that in view of the first and the second of eqs.(\ref{note2})
and of the first and the second of eqs.(\ref{note4}) the following
relations hold
\begin{eqnarray}
\C_a^\mu = -2 h_a^\mu \, , ~~~~ \h_a^\mu = -\frac{1}{2} C_a^\mu \, .
\label{new5}
\end{eqnarray}
Each of the choices in eq.(\ref{note1}) and (\ref{note3})
uniquely determines the set of ghosts and antighosts
of the model in the second order formalism, as explained in
Appendix \ref{appA}.
This in turn leads to the identification of two 
Hilbert spaces ${\cal H}_1$ and ${\cal H}_2$, both invariant
under the S-matrix.
The construction of ${\cal H}_{1,2}$ is also reported in Appendix \ref{appA}.

It is found that ${\cal H}_1$ is spanned by the four components
of $A^{'}_{a\mu}$ and by the transverse components of
$C_{a\mu}$, while ${\cal H}_2$ is spanned by the transverse
components of $A^{'}_{a\mu}$ and by the transverse components
of $\C_{a\mu}$. 
We identify the physical Hilbert space of the theory as the intersection
\begin{eqnarray}
{\cal H} = {\cal H}_1 \cap {\cal H}_2 \, .
\label{hphys}
\end{eqnarray}
In view of the relations in eq.(\ref{new5}) we conclude that
${\cal H}$ is spanned by the transverse components of $A'_{a\mu}$ only.

\medskip
In the $b^0_\mu$-sector there are no physical states, as proven in
Appendix \ref{appC}. Hence ${\cal H}$ is eq.(\ref{hphys})
represents the physical Hilbert space of the full minimal
embedding model.
This is consistent with the analysis of the off-shell cohomology
given in sect.~\ref{cohom}.

\section{Conclusions}
\label{sect4}

In this paper we have discussed a mechanism to generate
a mass for non-Abelian gauge bosons, based on a fixing
condition of an additional shift symmetry of a larger
model, into which massive Yang-Mills theory can be embedded.

In the embedding theory all the fields, besides the gauge
fields $A_\mu^a$ and their gauge ghosts
$\omega^a$, pair into BRST doublets together with their
shift ghosts partners.
Therefore these additional fields do not affect the cohomology
classes of the BRST differential $s$ under which the
embedding model is invariant. As a consequence, the cohomology
classes of $s$ in the FP-neutral sector 
(physical observables) reduce to local gauge-invariant
quantities constructed only from the field strength and its
covariant derivatives.

\medskip
We have shown that the embedding procedure fails to produce
a power-counting renormalizable theory when applied to the
field content of the ordinary Higgs model with a single
Higgs multiplet. 

Nevertheless such an analysis naturally suggests
the field content of a possible successful embedding of
massive Yang-Mills theory, which we have called
the minimal embedding model.
In this latter theory only one additional vector field $b_\mu$
appears in the bosonic sector together with the gauge fields
$A_\mu^a$.
\medskip

The minimal embedding model fulfills both the UV and the IR power-counting
conditions and is therefore  renormalizable by power-counting.
It also shows a simple cohomological structure in the extended
field sector. In this theory spontaneous gauge symmetry breaking
is not realized, as a consequence of the fixing condition 
in eq.(\ref{min2}).

We have discussed the spectrum of the 
classical action and we have analyzed which constraints
have to be imposed on the parameters $\lambda,\gamma$,
entering in eq.(\ref{min2}), in order to obtain a set of massive
non-Abelian gauge bosons.

In view of the cohomological structure of the relevant embedding
BRST differential, the physical observables
of the minimal embedding theory are given by the gauge-invariant local
quantities, constructed only from the field strength $F_{\mu\nu}^a$ and
its covariant derivatives.
The physical asymptotic states are the transverse polarizations
of the massive  fields $A^{'a}_\mu$.
The minimal embedding model could henceforth be regarded
as a candidate for a renormalizable model of massive Yang-Mills theory.

Many more checks and investigations
are needed in order to settle its status
as a field-theoretical description of pure massive
gauge bosons.
We only hint at some of them here.
The classical action in eq.(\ref{eq1_1})
depends on a parameter $\lambda$,
which can take values outside the region
$-1<\lambda<0$. In the limit $\lambda \rightarrow 0$
the Higgs model without quartic potential is formally recovered, 
while in the limit $\lambda \rightarrow -1$
the scalar field $\partial b^0$ acquires infinite mass.

The parameter $\lambda$ cannot be fixed on the basis of symmetry
arguments relying on the structure of the BRST differential.
The question of the dependence of the physical observables
on $\lambda$ is a subject deserving further study,
in order to explore the physical predictions of the model
and to compare them with those of other theories allowing for
massive gauge bosons.
Closely related to this issue, 
the inclusion of massive fermions
is also
 worthwhile to be analyzed, in order to shed light on possible
phenomenological applications of the minimal embedding formalism.

It may also happen that the model discussed in this paper
could suggest other yet unexplored possibilities to provide
a  field-theoretical framework for the description of
massive Yang-Mills theory.

\section*{Acknowledgments}

Useful discussions with and valuable comments 
from R.~Ferrari, D.~Maison and  P.~Weisz are gratefully acknowledged.

\appendix

\section{Physical states in the $A'_{a\mu}-b_{c\nu}$-sector}\label{appA}
In this Appendix we construct the physical Hilbert spaces
${\cal H}_{1,2}$ identified respectively by the choice
in eq.(\ref{note1}) (type-I second order formalism)
and in eq.(\ref{note3}) (type-II second order formalism).

Each choice uniquely defines the set of ghost fields of the model
in the second order formalism. 
Once the set of ghost fields is constructed, the physical states
can be selected as the non-trivial cohomology classes of the relevant
linearized BRST differential $\hat s$. 

The analysis carried out here applies at the classical level.
At higher orders the relevant BRST charge, acting on the asymptotic
states, can differ from the linearized BRST differential $\hat s$
by factors related to wave-function renormalizations and
possible mixings among the fields \cite{Becchi:bd}. 
The kernel of the proof
remains however unaffected and can be extended without
additional difficulties to the full renormalized minimal
embedding model.

\subsection{Type-I second order formalism}
In order to define the ghost content of the model
in the second-order formalism the first step is to reconstruct
the squares from eq.(\ref{note1}). For this purpose we use 
eq.(\ref{eq1_10}) to express $A'_{a\mu}$ as a function
of $A_{a\mu}$ and $b_{a\mu}$:
\begin{eqnarray}
A'_{a\mu} = A_{a\mu} - \lambda gm b_{a\mu} \, .
\label{appn1}
\end{eqnarray}
By substituting eq.(\ref{appn1}) into eq.(\ref{note1}) and
by using the relations
\begin{eqnarray}
&& C_a^\mu =  (\square + g^2m^2) b_a^\mu \, , \nonumber \\
&& \partial C_a = (\square + (gm)^2) \partial b \, ,
\label{note_e3}
\end{eqnarray}
(a consequence of the first of eqs.(\ref{note2})) we get
\footnote{For notational convenience we do not
write explicitly the sum over the index $a$.}
\begin{eqnarray}
S_1 & =& \int d^4x \, \left (
\frac{1}{2} A_{a\mu} \square A_a^\mu + \frac{1}{2} (\partial A_a)^2 
\right . 
\nonumber \\
&& \left . ~~~~~~~~
+ \frac{1}{2} ( \lambda C_{a\mu} - gm A_{a\mu} + \partial_\mu (\partial b_a))^2
\right . \nonumber \\
& & \left . ~~~~~~~~ + h_{a\mu} [  C_a^\mu - (\square + g^2m^2) b_a^\mu ]  
\right . \nonumber \\
& & \left . ~~~~~~~~
- \frac{1}{2} (\partial A_a + gm \partial b_a)^2 
\right ) \, .
\label{note_e11}
\end{eqnarray}
The terms in the first line are the bilinear contributions
from $-\frac{1}{4}(F_{\mu\nu}^a)^2$.
The structure of the ghost system is completely determined
by eq.(\ref{note_e11}). The relevant linearized
BRST transformations are 
\begin{eqnarray}
&& \hat s A'_{\mu a} = \partial_\mu \omega_a - \lambda gm \xi_{a\mu} \, , 
~~~~ \hat s \omega_a = 0 \, ,\nonumber \\
&& \hat s \bar \omega_a = \partial A_a + gm \partial b_a \, , \nonumber \\
&& \hat s \bar \theta_{a\mu} = h_{a\mu} \, , ~~~~ \hat s h_{a\mu} = 0 \, , 
~~~~ \hat s C_{a\mu} = \theta_{a\mu} \, , ~~~~ \hat s \theta_{a\mu} = 0 \, , \nonumber \\
&& \hat s \psi_{a\mu} = \lambda C_{a\mu} - gm A_{a\mu} + \partial_\mu \partial b_a \, , 
\nonumber \\
&& \hat s b_{a\mu} = \xi_{a\mu} \, , ~~~~ \hat s \xi_{a\mu} = 0 \, .
\label{note_e18}
\end{eqnarray}
The complete action $S_1$, including the ghost-dependent
terms, is
\begin{eqnarray}
S_1 & =& \int d^4x \, \left (
\frac{1}{2} A_{a\mu} \square A_a^\mu + \frac{1}{2} (\partial A_a)^2 
\right . 
\nonumber \\
&& \left . ~~~~~~~~
+ \frac{1}{2} ( \lambda C_{a\mu} - gm A_{a\mu} + \partial_\mu (\partial b_a))^2
+ h_{a\mu} [  C_a^\mu - (\square + g^2m^2) b_a^\mu ]  
\right . \nonumber \\
& & \left . ~~~~~~~~
- \frac{1}{2} (\partial A_a + gm \partial b_a)^2
\right . \nonumber \\
& & \left .  ~~~~~~~~ - \psi_{a\mu} [ \lambda \theta_a^\mu - gm \partial^\mu \omega_a
+ \partial^\mu (\partial \xi_a) ] 
- \bar \theta_{a\mu} [ \theta_a^\mu - (\square + g^2m^2) \xi_a^\mu]
\right . 
\nonumber \\
& & \left .  ~~~~~~~~ + \bar \omega_a ( \square \omega_a + gm \partial \xi_a) \right ) \, .
\label{note_e19}
\end{eqnarray}
The fields of the model are $A'_{a\mu}, h_{a\mu}, C_{a\mu}$.
Therefore we need to eliminate $b_{a\mu}$ in favour of $h_{a\mu}$ and
$C_{a\mu}$.
By using the first and the second of eqs.(\ref{note2}) we
obtain
\begin{eqnarray}
b_a^\mu = \frac{1}{g^2m^2}
\left [ C_a^\mu - \frac{1}{\lambda^2}
\left ( -2 h_a^\mu - \frac{(1+2\lambda)}{g^2m^2} \partial^\mu
\left ( \partial C_a + \frac{2}{(1+\lambda)^2} \partial h_a
\right ) \right ) \right ] \, ,
\label{note_e17}
\end{eqnarray}
valid for $\lambda \neq -1$.
By taking the $\hat s$-variation of both sides of eqs.(\ref{note_e17})
we get a constraint on $\xi^\mu_a$:
\begin{eqnarray}
\xi_a^\mu = \frac{1}{g^2m^2} \left [ \theta_a^\mu + 
\frac{1 + 2\lambda}{\lambda^2} \frac{1}{g^2m^2} \partial^\mu (\partial \theta_a) \right ] \, .
\label{note_e21}
\end{eqnarray}
This is valid for $\lambda \neq -1$. 
For  $\lambda \neq -1$ it can be proven
that it is equivalent to the constraint
\begin{eqnarray}
\hat s h_a^\mu = \lambda^2 \square \xi_a^\mu + (1 + 2 \lambda) \partial^\mu (\partial \xi_a) = 0 
\label{e21_bis}
\end{eqnarray}
The latter expression is however valid also for $\lambda = -1$.

\medskip
We can now move to the analysis of
\begin{eqnarray}
{\cal H}_1 = {\rm Ker } \ \hat s / {\rm Im } \ \hat s \, .
\label{appn5}
\end{eqnarray}
By imposing the equations of motion in the ghost sector
with the constraint in eq.(\ref{note_e21}) one finds
\begin{eqnarray}
\xi^\mu_a = \frac{1}{\lambda gm} \partial^\mu \omega_a \, , 
~~~~
\theta^\mu_a = \frac{1}{\lambda} gm \partial^\mu \omega_a \, ,
\label{appn6}
\end{eqnarray}
with
\begin{eqnarray}
\square \omega_a = 0 \, .
\label{appn7}
\end{eqnarray}
From eq.(\ref{note_e18}) the BRST transformations in the bosonic
sector become
\begin{eqnarray}
\hat s A'_{a\mu} = 0 \, , ~~~~~
\hat s C_{a\mu} = \frac{1}{\lambda} gm \partial_\mu \omega_a \, , ~~~~~
\hat s h_{a\mu} = 0 \, .
\label{appn8}
\end{eqnarray}
Moreover $h_{a\mu}$ is $\hat s$-exact (since $\hat s \psi_{a\mu} = h_{a\mu}$)
and $\bar \omega_a$ forms a doublet with $\partial A_a + gm \partial b_a$.
$\xi^\mu_a$ forms a doublet with $b^\mu_a$ given in
eq.(\ref{note_e17}).
We conclude from the above remarks and
from eq.(\ref{appn8}) that the space ${\cal H}_1$ in eq.(\ref{appn5})
is spanned by the four components of 
$A'_{a\mu}$ and by the transverse components of $C_{a\mu}$.

\subsection{Type-II second order formalism}

As before we start by reconstructing the squares from
eq.(\ref{note3}). We use eq.(\ref{appn1}) and the 
first of eqs.(\ref{note4}) to rewrite $S_2$ 
in eq.(\ref{note3}) as follows:
\begin{eqnarray}
S_2 & = & \int d^4x \, \Big (
+\frac{1}{2} A_{a\mu} \square A_a^\mu + \frac{1}{2} (\partial A_a)^2
\nonumber \\
& & ~~~~~~~~~
+ \h_{a\mu} (\C_a^\mu - \lambda^2 \square b_a^\mu -
(1 + 2\lambda) \partial^\mu (\partial b_a) )
\nonumber \\
& & ~~~~~~~~~
+ \frac{1}{2} \Big ( \frac{1}{\lambda} \C_a^\mu + 
\lambda (gm)^2 b_a^\mu - gm A_a^\mu -
\frac{1+\lambda}{\lambda} \partial^\mu (\partial b_a) \Big )^2
\nonumber \\
& & ~~~~~~~~~
- \frac{1}{2} (\partial A_a + gm \partial b_a)^2 \Big ) \, .
\label{note_t9}
\end{eqnarray}
Eq.(\ref{note_t9}) fixes the linearized BRST transformations
of the fields:
\begin{eqnarray}
&& \hat s A'_{a\mu} = \partial_\mu \omegaT_a - \lambda gm \xiT_{a\mu} \, , 
~~~~~ \hat s \omegaT_a = 0 \, , \nonumber \\
&& \hat s \baromegaT_a = \partial A_a + gm \partial b_a \, , \nonumber \\
&& \hat s \psiT_{a\mu} = \frac{1}{\lambda} \C_{a\mu}
+\lambda (gm)^2 b_{a\mu} - gm A_{a\mu}
-\frac{1+\lambda}{\lambda} \partial_\mu(\partial b_a) \, ,
\nonumber \\
&& \hat s \barthetaT_{a\mu} = \h_{a\mu} \, , ~~~~~~ \hat s \h_{a\mu} = 0 \, ,
\nonumber \\
&& \hat s \C_{a\mu} = \thetaT_{a\mu} \, , ~~~~~~ \hat s \thetaT_{a\mu} = 0 \, ,
\nonumber \\
&& \hat s b_{a\mu} = \xiT_{a\mu} \, , ~~~~~~ \hat s \xiT_{a\mu} = 0 \, .
\label{note_gh_tw2}
\end{eqnarray}
The complete action $S_2$, including the ghost-dependent terms, is
\begin{eqnarray}
S_2 & = & \int d^4x \, \Big (
+\frac{1}{2} A_{a\mu} \square A_a^\mu + \frac{1}{2} (\partial A_a)^2
\nonumber \\
& & ~~~~~~~~~
+ \h_{a\mu} (\C_a^\mu - \lambda^2 \square b_a^\mu -
(1 + 2\lambda) \partial^\mu (\partial b_a) )
\nonumber \\
& & ~~~~~~~~~
+ \frac{1}{2} \Big ( \frac{1}{\lambda} \C_a^\mu + 
\lambda (gm)^2 b_a^\mu - gm A_a^\mu -
\frac{1+\lambda}{\lambda} \partial^\mu (\partial b_a) \Big )^2
\nonumber \\
& & ~~~~~~~~~
- \frac{1}{2} (\partial A_a + gm \partial b_a)^2 
\nonumber \\
& & ~~~~~~~~~
- \psiT_{a\mu} (\frac{1}{\lambda} \thetaT_a^\mu + 
\lambda (gm)^2 \xiT_a^\mu - gm \partial^\mu \omegaT_a
-\frac{1+\lambda}{\lambda} \partial^\mu (\partial \xiT_a) )
\nonumber \\
& & ~~~~~~~~~
+\baromegaT_a (\square \omegaT_a+ gm \partial \xiT_a) 
\nonumber \\
& & ~~~~~~~~~
-\barthetaT_{a\mu} ( \thetaT_a^\mu 
-\lambda^2 \square \xiT_a^\mu - (1+2\lambda)\partial^\mu(\partial \xiT_a)) \Big ) \,.
\label{note_gh_tw1}
\end{eqnarray}
Again we eliminate $b_{a\mu}$ in favour of $\C_{a\mu}$ and
$\h_{a\mu}$. By using the first and the second of eqs.(\ref{note4})
we get
\begin{eqnarray}
b_a^\mu = -\frac{2}{(gm)^2}
\Big [ \h_a^\mu + \frac{1}{2\lambda^2}
\Big ( \C_a^\mu + \frac{(1 + 2 \lambda)}{(gm)^2}
\partial^\mu \Big ( \frac{1}{(1+\lambda)^2} \partial \C_a
+ 2 \partial \h_a \Big ) \Big ) \Big ] \, ,
\label{note_t14}
\end{eqnarray}
valid for $\lambda \neq -1$.
By taking the $\hat s$-variation of both sides of eqs.(\ref{note_t14})
we get a constraint on $\xiT^\mu_a$:
\begin{eqnarray}
\xiT_a^\mu = -\frac{2}{(gm)^2} \Big [
\frac{1}{2\lambda^2} \Big ( \thetaT_a^\mu 
+ \frac{1+2\lambda}{(1+\lambda)^2} \frac{1}{(gm)^2}
\partial^\mu (\partial \thetaT_a) \Big ) \Big ] \, .
\label{note_gn_tw18}
\end{eqnarray}
This condition is valid for $\lambda \neq -1$. 
For  $\lambda \neq -1$ 
it can be shown to be equivalent to the constraint
\begin{eqnarray}
\hat s \h_a^\mu = 
-\frac{1}{2} (\square + (gm)^2) \xiT^\mu_a = 0 \, .
\label{constr_tw}
\end{eqnarray}
The latter expression is however valid also for $\lambda = -1$.

The equations of motion in the ghost sector admit a solution
for $\xiT_{a\mu}$ in the same class as in eq.(\ref{appn6}),
given by
\begin{eqnarray}
\xiT_{a\mu}= -\frac{1}{gm} \partial_\mu \omegaT_a \, .
\label{tw10}
\end{eqnarray}
The constraint in eq.(\ref{constr_tw}) implies
\begin{eqnarray}
(\square + (gm)^2)\omegaT_a = 0 \, .
\label{tw11}
\end{eqnarray}
$\thetaT_{a\mu}$ is then given by 
\begin{eqnarray}
\thetaT_{a\mu} =  (1 + \lambda)^2 (gm) \partial_\mu \omegaT_a \, .
\label{tw12}
\end{eqnarray}
The linearized BRST transformations in the bosonic sector hence become
\begin{eqnarray}
&& \hat s A'_{a\mu} = (1+\lambda) \partial_\mu \omegaT_a \, , 
\nonumber \\
&& \hat s \C_{a\mu} = (1+\lambda)^2 gm \partial_\mu \omegaT_a \, , 
\nonumber \\
&& s \h_{a\mu} = 0 \,.
\label{tw13}
\end{eqnarray}
Moreover $\h_{a\mu}$ is $\hat s$-exact (since $\hat s \psiT_{a\mu} = \h_{a\mu}$)
and $\baromegaT_a$ forms a doublet with $\partial A_a + gm \partial b_a$.
$\xiT^a_\mu$ forms a doublet with $b^a_\mu$ in eq.(\ref{note_t14}).
\medskip
The space
\begin{eqnarray}
{\cal H}_2 = {\rm Ker \ } \hat s / {\rm Im \ } \hat s
\label{tw14}
\end{eqnarray}
is spanned by the transverse components of $A'_{a\mu}$
and of $\C_{a\mu}$,


\section{Physical states in the $b^0_\mu$-sector}\label{appC}

In the second order formalism 
 the
relevant quadratic part in the $b^0_\mu$-sector 
for the action given in eq.(\ref{eq1_1}), including
the ghost-dependent terms, is
\begin{eqnarray}
S_0 & = & \int d^4x \, 
\Big ( -\frac{1}{2} C^0_{\mu} C^{0\mu}
- C^0_\mu ( \lambda(\square + (gm)^2) b^{0\mu} + \partial^\mu(\partial b^0)
) \nonumber \\
& & ~~~~~~~~~~~~~
+ \psi^0_\mu ( \lambda (\square + (gm)^2) \xi^{0\mu} + \partial^\mu (\partial \xi^0) )
 \Big ) \, .
\label{note_eb1}
\end{eqnarray}

The equations of motion in the bosonic sector are
\begin{eqnarray}
\frac{\delta S_0}{\delta C^0_\mu} & = &
- C^0_\mu - \lambda (\square + (gm)^2) b^0_\mu - \partial_\mu (\partial b^0)
= 0 \, , \nonumber \\
\frac{\delta S_0}{\delta b^0_\mu} & = &
-\lambda (\square + (gm)^2) C^0_\mu - \partial_\mu (\partial C^0) = 0 \, .
\label{note_eb3}
\end{eqnarray}
From the first of eqs.(\ref{note_eb3}) we get
\begin{eqnarray}
 C^0_\mu =  - \lambda (\square + (gm)^2) b^0_\mu - \partial_\mu (\partial b^0) \, .
\label{note_eb4}
\end{eqnarray}
From the second of eqs.(\ref{note_eb3}) we get
\begin{eqnarray}
\lambda (\square + (gm)^2) C^0_\mu + \partial_\mu (\partial C^0) = 0 \, .
\label{note_eb5}
\end{eqnarray}
By taking the divergence of eq.(\ref{note_eb5}) we find
\begin{eqnarray}
(1 + \lambda) \square \partial C^0 + \lambda (gm)^2 \partial C^0 = 0 \, .
\label{note_eb6}
\end{eqnarray}
The scalar $\partial C^0$ has mass $p^2 = \frac{\lambda}{1 + \lambda} (gm)^2$.
It goes to infinity for $\lambda \rightarrow -1$.

\medskip
The relevant linearized BRST transformations are
\begin{eqnarray}
\hat s \psi^0_\mu = C^0_\mu \, , ~~~~ \hat s C^0_\mu = 0 \, , 
~~~~
\hat s b^0_\mu = \xi^0_\mu \, , ~~~~ \hat s \xi^0_\mu = 0 \, .
\label{note_eb2}
\end{eqnarray}
In view of eq.(\ref{note_eb2}) we see that there are no
additional physical particles in the $b^0_\mu$-sector.

We wish to make a comment on the limit $\lambda \rightarrow -1$.
In this limit we see from the equation of motion of $\psi^0_\mu$ that
\begin{eqnarray}
\partial \xi_0 = 0 \, .
\label{app20}
\end{eqnarray}
Therefore the scalar $\partial b_0$ would be physical
(since $\hat s (\partial b_0)= \partial \xi_0 = 0$)
However, in the limit $\lambda \rightarrow -1$ 
we get from eq.(\ref{note_eb4}) that
\begin{eqnarray}
\partial b^0 = \frac{1}{(gm)^2} \partial C^0 \, .
\label{app21}
\end{eqnarray}
The scalar $\partial C^0$ in turn becomes infinitely massive
in the limit $\lambda \rightarrow -1$ and therefore it disappears
from the physical spectrum.

\section{UV and IR degrees of the minimal embedding model}\label{appB}

\
\vskip 0.3 truecm

\begin{center}
\begin{tabular}{|c|c|c|}
\hline
 Field & UV dim. ($\delta$) & IR dim. ($\rho$) \cr
\hline
$A_{a\mu}$ & 1 & 1 \cr
\hline
$b_{0\mu}$ & 0 & 2 \cr
\hline
$b_{a\mu}$ & 0 & 1 \cr
\hline
\hline
\hline
$\psi_{0\mu}$ & 1 & 2 \cr
\hline
$\psi_{a\mu}$ & 1 & 2 \cr
\hline
$\xi_{0\mu}$ & 1 & 2 \cr
\hline
$\xi_{a\mu}$ & 1 & 2 \cr
\hline
$\bar \omega^a$ & 1 & 1 \cr
\hline
$\omega^a$ & 1 & 1 \cr
\hline
\end{tabular}
\end{center}
\begin{center}
Table 1 - UV and IR assignments for the fields of the minimal embedding model.
\end{center}

\end{document}